# Magnetization precession induced by quasi-transverse picosecond strain pulses in (311) ferromagnetic (Ga,Mn)As

M. Bombeck[1], J.V. Jäger[1], A.V. Scherbakov[2], T. Linnik[3], D.R. Yakovlev[1,2], X. Liu[4], J.K. Furdyna[4], A.V. Akimov[2,5] and M.Bayer[1,2]

[1]*Experimentelle Physik 2, Technische Universität Dortmund, D-44227 Dortmund, Germany*

[2]*Ioffe Physical-Technical Institute, Russian Academy of Sciences, 194021 St. Petersburg, Russia*

[3]*Department of Theoretical Physics, V.E. Lashkaryov Institute of Semiconductor Physics, 03028 Kyiv, Ukraine*

[4]*Department of Physics, University of Notre Dame, Notre Dame, Indiana 46556, USA*

[5]*School of Physics and Astronomy, University of Nottingham, Nottingham NG7 2RD, UK*

Abstract

Quasi-longitudinal and quasi-transverse picosecond strain pulses injected into a ferromagnetic (311) (Ga,Mn)As film induce dynamical shear strain in the film, thereby modulating the magnetic anisotropy and inducing resonant precession of the magnetization at a frequency ~10 GHz. The modulation of the out-of-plane magnetization component by the quasi-transverse strain reaches amplitudes as large as 10% of the equilibrium magnetization. Our theoretical analysis is in good agreement with the observed results, thus providing a strategy for ultrafast magnetization control in ferromagnetic films by strain pulses.

The inverse magnetostrictive effect (the Villari effect) in ferromagnetic materials is widely used in sensors and acousto-magnetic transducers operated at frequencies up to hundreds of kHz. A further increase of the manipulation speed, extending it to gigahertz (GHz) or sub-terahertz (THz) frequency, is desirable, because it would open new prospects for high-frequency electronics, computing, and information processing. Although this is a highly challenging task, several important experimental and theoretical advances have recently been made in this direction by demonstrating modulation of magnetization through injection of high-frequency acoustic waves into nanometer ferromagnetic films. Specifically, surface acoustics waves with frequencies ~100 MHz were applied to 10 nm Co films for this purpose [1]; coherent precession of magnetization and excitation of spin waves at ~10 GHz frequency were achieved by longitudinal picosecond strain pulses in (Ga,Mn)As [2,3] and Ni [4] films; and comprehensive theoretical analysis of the magnetization dynamics in presence of picosecond strain pulses was elaborated [5].

For practical applications of GHz and sub-THz magnetostrictive phenomena it is mandatory to find materials and conditions where the magnetization modulation occurs with significant amplitudes at these high frequencies. The inherent dependence of magnetic properties on strain provides a powerful tool for ultrafast modulation of a ferromagnetic nanostructure by picosecond strain pulses. Consider the most common case of a cubic ferromagnetic film. At zero or small external magnetic field, the magnetization is oriented along one of the easy magnetization axes (commonly in the layer plane) as result of magnetic anisotropy. This direction of magnetization may be tilted from its equilibrium position by applying external stress, but due to symmetry properties this will only occur if the strain tensor of the film has non-zero shear (i.e. non-diagonal) elements. In static experiments a strong magnetic response to elastic perturbations takes place when the strain is induced by a force applied in the film plane at a finite angle relative to the easy axis of magnetization [6, 7].

These considerations can be exploited for magnetization modulation and control by acoustic waves or strain pulses which generate shear strain in the ferromagnetic film. This can be achieved using surface acoustic waves, which are, however, typically limited to frequencies of ~1 GHz, so that higher frequency magneto-acoustic resonances cannot be reached [8]. As an alternative, one could use picosecond strain pulses, which involve frequencies up to ~100 GHz to induce resonant oscillations of the magnetic system at high frequencies [2-5]. However, the corresponding experiments performed so far involved purely longitudinal, i.e., purely compressive strain pulses propagating perpendicular to the film plane. Then the modulation of magnetization can only be obtained by applying additionally an external out-of-plane magnetic field [2,4].

Here we show that exploiting transverse acoustic modes, which generate dynamical shear strain, enforce an out-of-plane modulation, thereby increasing the efficiency of ultrafast magnetization control significantly. In detail, we study a crystalline magnetic film grown on a low-symmetry (311) crystallographic plane. In such a sample, a strain pulse generated by standard picosecond acoustic techniques [9] is given by a sequence of a quasi-longitudinal acoustic (QLA) wavepacket followed by a quasi-transverse acoustic (QTA) wavepacket [10,11]. These acoustic excitations propagating into the ferromagnetic film cause dynamically elastic shear impacts onto the magnetic system. We show experimentally that both the QLA and the QTA pulse induce a magnetization modulation, which for the out-of-plane component of the magnetization can becomes as large as 10% of its absolute value $M_0$ when the QTA pulse hits the film. Such a large change in magnetization, **M**, exceeds several times previously reported values in the GHz frequency range. Our theoretical analysis shows the way how to obtain optimum conditions for high-amplitude modulation of magnetization in ferromagnetic films by transverse picosecond strain pulses.

The sample used in the experiment is a ferromagnetic (311) (Ga,Mn)As film with a thickness $d$=85 nm grown by low-temperature molecular beam epitaxy (MBE) on a (311) semi-insulating GaAs substrate with 100-μm thickness [12]. The experiments are performed in a cryostat with a superconducting magnet at temperature $T$ = 6 K, well below the Curie temperature $T_c$ = 60 K of the film. The magnetic field **H** is applied in the plane of the (Ga,Mn)As film, forming an angle $\varphi_H$ with the [$\bar{2}33$] crystallographic axis. The geometry of the experiment, including the definition of the basis axes ($x$, $y$ and $z$), are shown in Fig. 1(a). The two easy magnetization axes in (Ga,Mn)As are along the [001] and [010] crystallographic directions of the cubic film [see insert in Fig. 1 (a)], none of which lies in the (311) plane of the film. Applying an in-plane field **H** turns the equilibrium magnetization **M**$_0$ towards the (311) plane, and for $\mu_0 H >$ 0.25 T ($\mu_0$ is vacuum permeability) **M**$_0$ may with high accuracy be considered to fall into the film, i.e. $M_z \ll M_0$. The magnetic properties of the sample, including its anisotropy parameters and magnetization curves, were studied in detail in earlier work [12].

The picosecond strain pulses were generated by pulsed optical pump excitation of a 100 nm Al film deposited on the GaAs substrate, opposite to the (Ga,Mn)As layer (for details see [13]). The strain pulses injected in this way into the GaAs substrate are propagating along the [311] crystallographic axis. Solution of the elastic equation for propagation of acoustic waves along this low-symmetry direction indicates that two acoustic modes are generated, QLA and QTA, with sound velocities of $s_{LA} = 5.1 \times 10^3$ m/s and $s_{TA} = 2.9 \times 10^3$ m/s, respectively [14]. In the basis of the axes shown in Fig. 1(a) the QLA and QTA strain pulses have the following strain tensor amplitude components: $\varepsilon_{zz}^{LA} = 0.918\varepsilon_0$, $\varepsilon_{xz}^{LA} = 0.077\varepsilon_0$, $\varepsilon_{yz}^{LA} = 0$, $\varepsilon_{zz}^{TA} = 0.067\varepsilon_0$, $\varepsilon_{xz}^{TA} = -0.2\varepsilon_0$, and $\varepsilon_{yz}^{TA} = 0$. Here the factor $\varepsilon_0$ is the amplitude of the compressive (i.e. longitudinal) strain pulse excited in an isotropic medium: $\varepsilon_0 = k_w W$, where coefficient $k_w \sim 10^{-4}$ cm$^2$/mJ and $W$ is the excitation density on the Al film [15]. The temporal profiles $\varepsilon_{ij}^q(t)$ for the strain pulses ($q$ gives their polarization: QLA or QTA) injected into the GaAs substrate are shown in Figs. 1 (b) and (c) for $W=3$ mJ/cm$^2$. The spectra of these QLA and QTA acoustic wave packets obtained by fast Fourier transformation of $\varepsilon_{ij}^q(t)$ are shown in the insets.

The QLA and QTA strain pulses propagate through the GaAs substrate, eventually reaching the (Ga,Mn)As layer after $t_{LA} \approx 19.6$ ns and $t_{TA} \approx 34.5$ ns, respectively. At these moments the strain pulses perturb the magnetic anisotropy of the layer, leading to ultrafast modulation of the magnetization. The temporal evolution of **M**(t) is monitored by measuring the angle of Kerr rotation $\Delta\psi(t)$ of a weak, linearly polarized optical probe pulse taken from the same femtosecond laser as the pump and hitting the sample normal to the film plane [13]. For $\mu_0 H > 0.25$ T, when **M**$_0$ is in the plane of the film, the elasto-optical contribution to Kerr rotation from circular dichroism is negligible [16]. By studying the Kerr signals for different linear polarizations of the probe beam, we verified that linear dichroism [3] does not contribute to $\Delta\psi(t)$. Thus measurement of $\Delta\psi(t)$ provides directly the temporal evolution of the $z$-component of the magnetization, $M_z(t)$, with the relative changes $\Delta M_z(t)$ given by $\Delta M_z(t)/M_0 = K\Delta\psi(t)$, where the coefficient $K = 140$ rad$^{-1}$ was obtained from Kerr rotation calibration measurements in the Faraday geometry (**H**∥**z**) in the absence of strain pulses.

The solid lines in Fig. 2 show the signals $\Delta M_z(t)/M_0$ measured at $\varphi_H=0$ (**H**∥[$\bar{2}$33]) using a low pump excitation density $W=3$ mJ/cm$^2$ for three values of $H$. Nonzero $\Delta M_z(t)/M_0$ is detected in the time intervals when the QLA and QTA strain pulses are propagating in the (Ga,Mn)As film. The signals are oscillatory with a period that decreases with increasing magnetic field. The signals decay during several hundreds of picoseconds. Note that oscillation period and decay time of $\Delta M_z(t)/M_0$ are similar for QTA and QLA at a given $H$.

The magnetic field dependences for the root mean square (RMS) amplitude $\overline{M}_z$ averaged over a 0.3 ns time interval are shown in Fig. 3(a). These dependences show broad maxima at $\mu_0 H_m^{TA}$=0.4 T and $\mu_0 H_m^{LA}$=0.7 T for the signals induced by the QLA and QTA modes, respectively. Figure 3(b) compares $\Delta M_z(t)/M_0$ of the QTA pulse measured at $\varphi_H$=0 and $\varphi_H$=π/2. It is clearly seen that the amplitude of the signal measured at $\varphi_H$=π/2 is drastically weaker than at $\varphi_H$=0.

The observation of oscillatory signals $\Delta M_z(t)/M_0$ induced by QTA strain pulses is the main experimental result of the present work, and below we discuss results of corresponding calculations (for details of the underlying theoretical formulation see [17]). From earlier theoretical studies [5] it is known that three factors govern $\Delta M_z(t)/M_0$: (*i*) the angle by which the strain tilts the axis of magnetization precession from its equilibrium orientation **M**$_0$, which is equivalent to the generation of a torque for precession of **M**; (*ii*) the degree to which synchronization is fulfilled in the spin-phonon interaction (in analogy with wave vector selection rules); and (*iii*) the existence of acoustic modes in the strain pulse spectrum resonant with magnetic excitations.

The symmetry of the used experimental geometry [Fig. 1 (a)], for $\varphi_H$=0 and at $\mu_0 H$>0.25 T results in time-dependent tensor components $\varepsilon_{zz}^q$ and $\varepsilon_{xz}^q$ launching the magnetization precession around an axis, that is tilted out of the film plane purely vertically. The tilting angle, which determines factor (*i*), can be described as the change $\Delta\theta$ of the polar angle $\theta$ between the [311] direction and the equilibrium orientation of **M**$_0$:

$$\Delta\theta_q = A\varepsilon_{zz}^q + B\varepsilon_{xz}^q \qquad (1)$$

The coefficients *A* and *B* in Eq. (1) are determined by the magnetic anisotropy parameters and show complex dependences on the equilibrium direction of the magnetization **M**$_0$ [5]. As an example for their magnitudes, numerical calculations for $\mu_0 H$=0.7 T give *A*=18 rad and *B*=82 rad. For strain pulses propagating along the [311] direction we thereby get as maximum values for the tilt angles $\Delta\theta_{LA} = 23\varepsilon_0$ and $\Delta\theta_{TA} = -15\varepsilon_0$. Note that $\Delta\theta_q$ plays the most important role in building up the precession amplitude $\bar{M}_z$ when the other two other factors (*ii*) and (*iii*), are close to their optimal values. Only then the maximum deviations $\Delta M_z(t) \sim -M_0 \Delta \theta_q$.

The synchronization condition (*ii*) is at an optimum when the time for a round trip of the acoustic phonon forward and backward through the magnetic film is close to the period of a magnetic precession [3, 5]. More precisely, the following equation should be fulfilled:

$$f_H = 0.37 s_q d^{-1}, \tag{2}$$

where $f_H$ is the $H$ dependent frequency of precession of **M** [see inset of Fig. 3 (a)].

The last factor (*iii*) which influences $\Delta M_z(t)$ is related to the spectral content of the acoustic strain pulses shown in the insets of Figs. 1 (b) and (c) for the QLA and QTA modes, respectively. These broad spectra are for both modes centered in the GHz frequency range of magnetization precession, making picosecond strain pulses an efficient tool for high-frequency resonant manipulations of **M**.

All three factors (*i*), (*ii*) and (*iii*) play an important role for the complicated trajectory of **M**(*t*) described by the Landau-Lifshitz equation. The dashed curves in Fig. 2 show the results of numerical calculations of $\Delta M_z(t)$ without taking into account the decay of precession. Most importantly the positions of maxima and minima of $\Delta M_z(t)$ occur at the same times in the calculated and measured traces, and the buildup of the oscillation amplitude at early times is also in excellent quantitative agreement. The calculated magnitude of $\Delta \theta_{LA}$ is close that of $\Delta \theta_{TA}$, and the experimental values of $\bar{M}_z$ for QLA and QTA [see Fig. 3(a)] have similar magnitudes for those values of $H$ when synchronization condition is closely fulfilled, both in experiment and theory. The measured $H_m^q$ [i.e., the fields at which $\bar{M}_z$ are maximum for the modes $q$, as illustrated in Fig. 3 (a)] are in good agreement with the values of $H$ [shown in Fig. 3 (a) by vertical arrows] corresponding to conditions of optimum synchronization according to Eq.(2). The calculated dependence of $\Delta M_z(t)/M_0$ on the angle $\varphi_H$ explains the large difference between the amplitudes of the signals measured at $\varphi_H=0$ and $\varphi_H=\pi/2$, as seen in Fig. 3(c). Qualitatively, the $[0\bar{1}1]$ direction ($\varphi_H=\pi/2$) is a specific direction, along which **M** is insensitive to strain: if **M** points along $[0\bar{1}1]$, the coefficients $A$ and $B$ in Eq. (1) decrease significantly relative to the $\varphi_H=0$ case, so that the tilt angle $\Delta\theta$ becomes negligible. There are several directions for which such reduction takes place, they are determined by the structural symmetry as is detailed in Ref. [5].

Therefore the theoretical model contributes to understanding and designing ultrafast modulation of magnetization for the most common case of its orientation in equilibrium **M**$_0$ in the film plane.

Next we turn to the measured dependence $\Delta M_z(t)/M_0$ on the pump excitation density $W$ in Fig. 4 for the QTA (a) and QLA (b) modes. The temporal evolution of $\Delta M_z(t)/M_0$ for the QTA mode remains unchanged with increasing excitation density $W$ [Fig. 4(a)] and, as shown in the inset of Fig. 4 (a), the amplitude $\overline{M}_z^{TA}$ increases linearly with $W$. A different behavior is observed for the QLA mode [see Fig. 4(b)]. Here the temporal evolution $\Delta M_z(t)/M_0$ stays the same only up to $W$=8 mJ/cm$^2$, but for higher $W$ strong non-linear effects start to become significant: the arrival time of the QLA strain pulse decreases, the duration of the response to QLA increases, and the value of $\overline{M}_z^{LA}$ tends to saturate with increasing $W$. These features of the QLA-induced signal are result of nonlinear propagation of the high-amplitude compressive acoustic waves in the GaAs substrate [15, 18]. For shear waves corresponding studies have not yet been performed, a detailed discussion of the nonlinear effects is beyond the scope of the present work. However, most importantly the linearity of $\overline{M}_z^{TA}$ with $W$ for the QTA mode results in values of $\Delta M_z(t)/M_0$ up to 10% at $W$=52 mJ/cm$^2$ (and potentially increases further for larger $\varepsilon_0$), while for the QLA mode $\Delta M_z(t)/M_0$ tends to saturate at ~2% for this value of $W$ [compare Figs. 4(a) and 4(b)]. The magnitude of the magnetization tilt is limited only by the acoustic energy in the QTA mode, i.e. by the laser power of our setup. A source capable of delivering stronger QTA waves may achieve full magnetization switching based on the demonstrated technique.

In conclusion, we have observed ultrafast resonant magnetization precession induced by quasi-transverse strain pulses which cause a shear perturbation acting on the magnetization lying in the plane of the ferromagnetic film. For low excitation densities the precession amplitudes of the signals induced by QTA and QLA strain pulses have similar values. Considering that only ~6% of the incident elastic energy is funneled into the QTA mode, we may conclude that QTA waves are considerably more efficient in inducing a magnetization modulation compared to QLA waves. The amplitude of precession induced by the QTA pulses increases linearly with strain amplitude, with $\Delta M_z(t)/M_0$ reaching a value of ~10% of the absolute equilibrium value limited only by technical constraints so that complete magnetization switching by QTA modes seems possible. These experiments, along with guidance by our theoretical analysis, provide a new instrument for ultrafast magnetization control using transverse acoustic pulses, and therefore open prospects for manipulating the magnetization in thin ferromagnetic films without applying an external magnetic field.

We acknowledge Boris Glavin and Vitalyi Gusev for fruitful discussions. The work was supported by the Deutsche Forschungsgemeinschaft (BA 1549/14-1), the Russian Foundation for Basic Research (11-02-00802), the Russian Academy of Science, and the US National Science Foundation grant DMR-1005851.

**Figure captions**

Figure 1

(a) The geometry of the experiment. The inset shows the directions of the two easy axes of magnetization, [001] and [010]. (b) Temporal evolutions of the tensor components $\varepsilon_{zz}$ and $\varepsilon_{xz}$ propagating along the [311] axis in GaAs, calculated for the QLA strain pulse generated in the Al film using an pump excitation energy density of 3 mJ/cm². The inset shows the spectral density of the acoustic wavepacket. (c) Same as (b), but for the QTA polarization. The calculations of these temporal evolutions take into account the acoustic mismatch at the Al/GaAs interface resulting in multiple peaks in the temporal and spectral profiles.

Figure 2

Time evolutions of the z-component of the magnetization modulation induced by QLA and QTA picosecond strain pulses measured for $W$=3mJ/cm² at different magnetic fields with **H**∥[$\bar{2}$33] ($\varphi_H$=0). The dotted curves show numerical calculations for zero damping. $t$=0 corresponds to the time of impact of the pump excitation pulse on the Al film.

Figure 3

(a) Magnetic field dependences of the RMS amplitudes of the measured Kerr rotation signals induced by QTA and QLA strain pulses. Vertical arrows indicate the calculated values of $H$ at which the synchronization condition for excitation of precession is fulfilled. The inset shows the calculated precession frequencies (solid line) and the precession frequencies obtained by fast Fourier transformation of the measured $\Delta M_z(t)$ (symbols) as function of applied magnetic field. (b) Temporal evolution of the z-component of the magnetization measured for the magnetic field applied along [$\bar{2}$33] ($\varphi_H$=0) and along [0$\bar{1}$1] ($\varphi_H$=π/2).

Figure 4

Temporal evolution of the z-component of the magnetization induced by QTA (a) and QLA (b) strain pulses for $\mu_0 H$=0.5 T (with **H**∥[$\bar{2}$33]) at different excitation power densities $W$. The inset in (a) shows the RMS amplitude of the measured signals induced by QTA pulses as function of $W$.

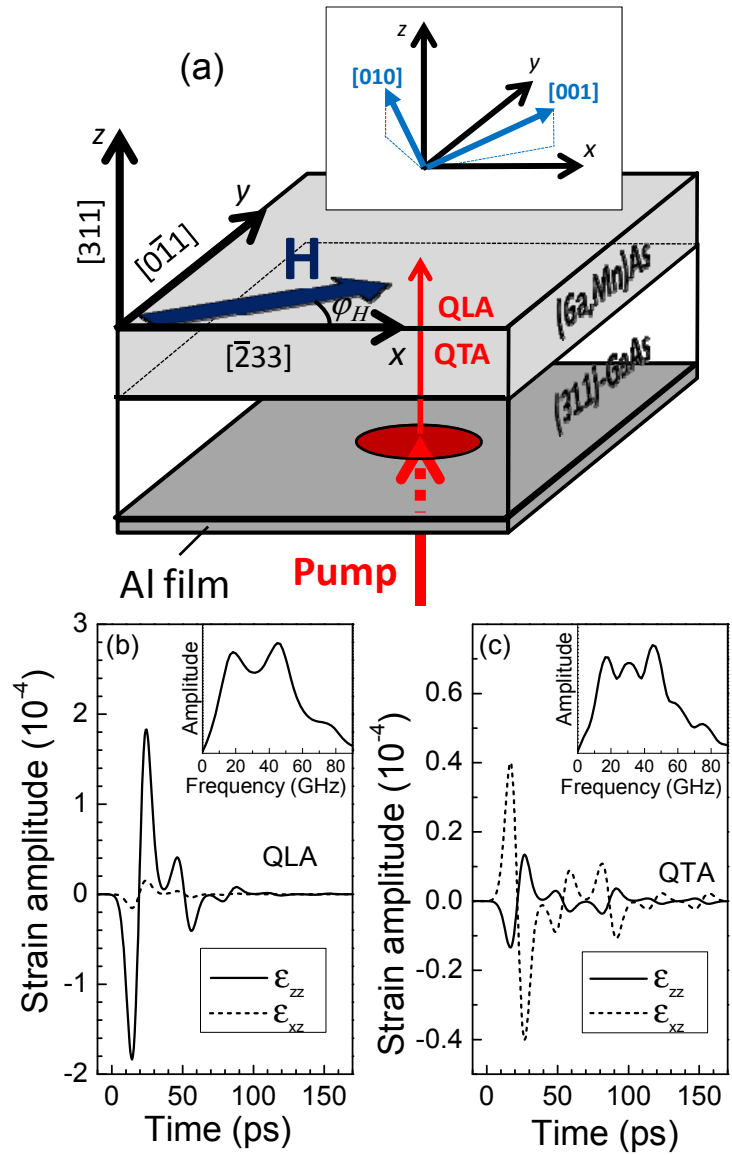

**Figure 1.** M. Bombeck et al. *"Magnetization precession induced by quasi-transverse picosecond strain pulses…"*

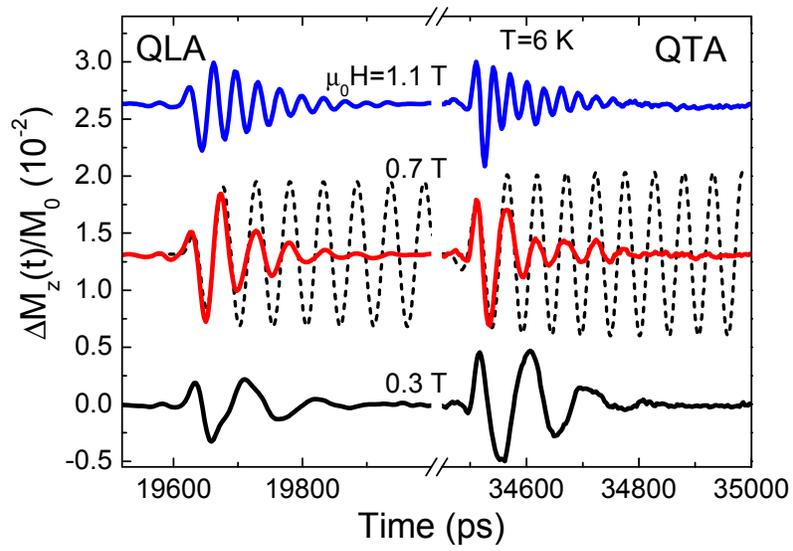

**Figure 2.** M. Bombeck et al. *"Magnetization precession induced by quasi-transverse picosecond strain pulses…"*

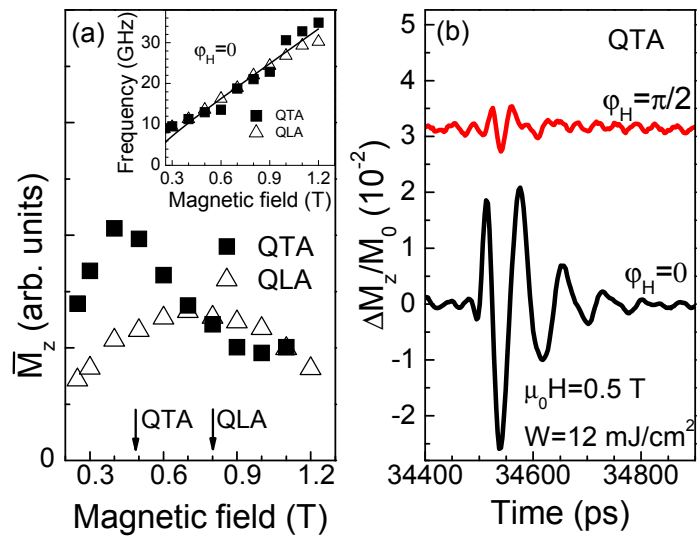

**Figure 3.** M. Bombeck et al. *"Magnetization precession induced by quasi-transverse picosecond strain pulses…"*

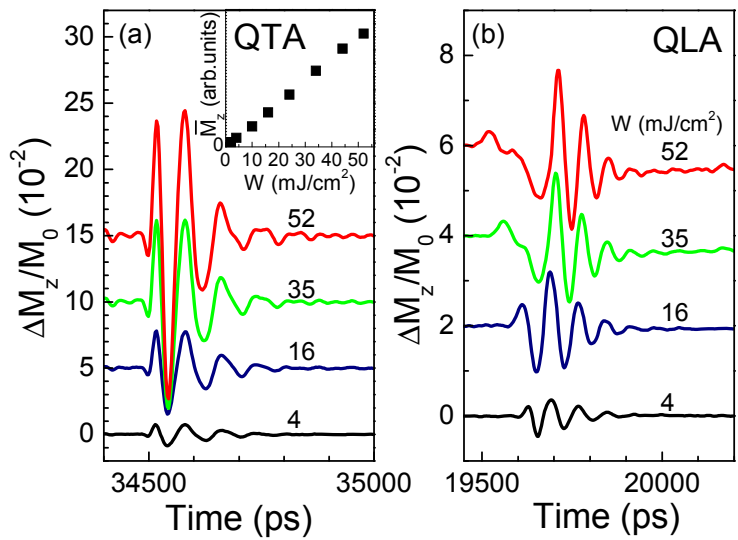

**Figure 4.** M. Bombeck et al. *"Magnetization precession induced by quasi-transverse picosecond strain pulses…"*